# Periodically-Poled Silicon

Nick K. Hon, Kevin K. Tsia, Daniel R. Solli, and Bahram Jalali[a]

*Electrical Engineering Department*

*University of California, Los Angeles, CA, 90095–1594*

Abstract

We propose a new class of photonic devices based on periodic stress fields in silicon that enable second-order nonlinearity as well as quasi-phase matching. Periodically-poled silicon (PePSi) adds the periodic poling capability to silicon photonics, and allows the excellent crystal quality and advanced manufacturing capabilities of silicon to be harnessed for devices based on second-order nonlinear effects. As an example of the utility of the PePSi technology, we present simulations showing that mid-wave infrared radiation can be efficiently generated through difference frequency generation from near-infrared with a conversion efficiency of 50%. This technology can also be implemented with piezoelectric material, which offers the capability to dynamically control the $\chi^{(2)}$ nonlinearity.

---

[a] Electronic mail: jalali@ucla.edu

As a centrosymmetric crystal, bulk silicon lacks second-order optical nonlinearity – a foundational component of nonlinear optics. Hence, its lowest-order nonlinearity originates from the third-order susceptibility $\chi^{(3)}$, which gives rise to the Raman and Kerr effects [1]. However, silicon's crystal symmetry can be broken to create second-order nonlinearity by applying a dc-electric field [2], operating at interfaces where the crystal symmetry is interrupted [3], and applying mechanical stress (or equivalently strain) in the material [4,5].

An intriguing property of parametric $\chi^{(2)}$ processes is the possibility of achieving quasi-phase matching (QPM) by periodic poling – a technique that enhances the efficiency of nonlinear interactions. Unfortunately, conventional poling processes, such as those used for lithium niobate and nonlinear polymers, do not apply to silicon because it lacks a dipole moment in its native form. Here, we propose an approach for realizing, what is in effect, periodically-poled silicon (PePSi), a new technology for efficient second-order nonlinear processes. We achieve this by creating alternating stress fields along a silicon waveguide using a periodic arrangement of stressed thin films. We show that this structure creates appreciable $\chi^{(2)}$ and, simultaneously, achieves QPM. Via comprehensive numerical simulations that include two-photon absorption (TPA), as well as free-carrier plasma effects and its wavelength dependence, we show that PePSi can be used for efficient mid-wave infrared (MWIR) generation. Such a capability has many applications including remote sensing of chemical and biological agents and environmental monitoring [6]. Piezoelectric thin films have previously been used to dynamically control phase matching in silicon wavelength converters that operate based on the $\chi^{(3)}$ nonlinear response [7]. PePSi can also be combined with piezoelectric layers, an powerful technology that offers the ability to dynamically control $\chi^{(2)}$ in silicon using intelligent electronic circuitry.

MWIR generation in silicon can be implemented by first-order stimulated Raman scattering (SRS) pumped at shorter MWIR wavelengths [8], or by cascaded SRS pumped at near infrared (NIR) wavelengths [9,10]. In contrast, our MWIR generation approach using PePSi is a single-step conversion from NIR via difference frequency generation (DFG). This is more favorable in terms of the wide-availability of the NIR pump sources and also more efficient in terms of circumventing the cumulative TPA and associated free-carrier absorption (FCA) that occurs in the cascaded SRS approach.

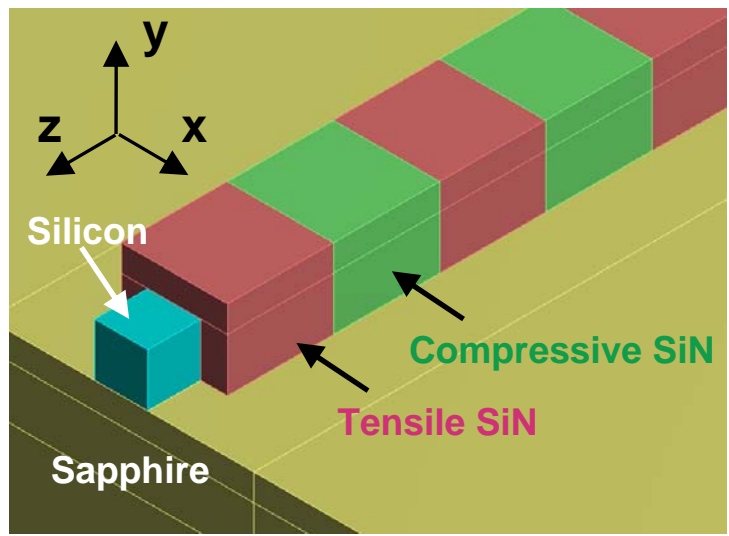

Fig. 1. (Color online) An example of PePSi waveguide formed by covering a silicon channel waveguide with two types of silicon nitride (SiN) stressed films periodically along the waveguide: one induces compressive stress, whereas the other one produces tensile stress.

The PePSi structure considered here is a channel waveguide integrated with two types of silicon nitride (SiN) stressed films: one with tensile stress and another with compressive stress, periodically deposited along the waveguide, as shown in Fig. 1. Hence, the sign of stress induced within the silicon core alternates along the waveguide and results in alternating dipole regions – a new type of periodically-poled structure.

SiN is chosen because it achieves high film stress [11] and is compatible with complementary metal-oxide-semiconductor processing. In addition, the force in SiN films can be readily tailored from compressive to tensile stress depending on the deposition conditions [11]. It thus permits flexible engineering of stress, and hence $\chi^{(2)}$, in silicon. In practice, the design shown in Fig. 1 can be realized by two different SiN deposition steps, giving rise to alternating stresses along the waveguide. For MWIR applications, silicon-on-insulator (SOI) is not the desirable platform because of the high losses of silicon dioxide at MWIR wavelengths. Instead, silicon-on-sapphire is employed here as sapphire is transparent to MWIR.

The PePSi waveguide we consider has the silicon core dimensions of 2 μm × 2 μm, which supports optical waveguiding in both MWIR and NIR regimes. We assume the in-plane stresses in the two different SiN films (with 1 μm thickness) to be +1 GPa (*tensile*) and –500 MPa (*compressive*) [11]. The SiN stressed film period is designed to be 8 μm in order to phase-match the interacting waves in the DFG process for MWIR generation. To estimate the stress-induced $\chi^{(2)}$ in the PePSi waveguide, we simulated the stress distribution in the waveguide by a three-dimensional finite-element analysis package (ANSYS). In principle, all the stress components (i.e., the normal and shear stresses in all directions) should be considered for evaluating the stress-induced $\chi^{(2)}$ because it is the anisotropy of the stress responsible for breaking the original crystal symmetry. Nevertheless, it is found that, compared to all other stress components, the *y*-component normal stress ($\sigma_{yy}$) gives rise to a highly uniform stress distribution in the silicon core with a considerably higher average stress magnitude in the present PePSi waveguide design. Hence, it is conceivable that stress-induced $\chi^{(2)}$ here is dominated by $\sigma_{yy}$. Simulated cross-sectional $\sigma_{yy}$ distributions in the tensile and compressive SiN cladding regions are shown in Fig. 2 (a) and (b), respectively. In one half of the period, when the SiN film exhibits tensile stress that deforms the waveguide, a compressive stress field is induced and confined inside the waveguide core (Fig. 2(a)) in order to counteract the deformation under elastic equilibrium. Conversely, in another half of the period, the compressive SiN cladding induces tensile stress within the silicon core (Fig. 2(b)). In addition, the silicon core, covered by the conformal SiN stressed cladding, displays a stress distribution, and hence the stress-induced $\chi^{(2)}$ distribution, with good uniformity (Fig. 2(a) – (b)). This feature is important to ensure efficient $\chi^{(2)}$ interaction of the optical modes within the waveguide. The picture of the periodic poling becomes more appealing when we observe the average stress ($\sigma_{yy}$) along the waveguide. As illustrated in Fig. 2(c), the periodic oscillation with –500 MPa peak average compressive stress and +200 MPa peak average tensile stress is evident.

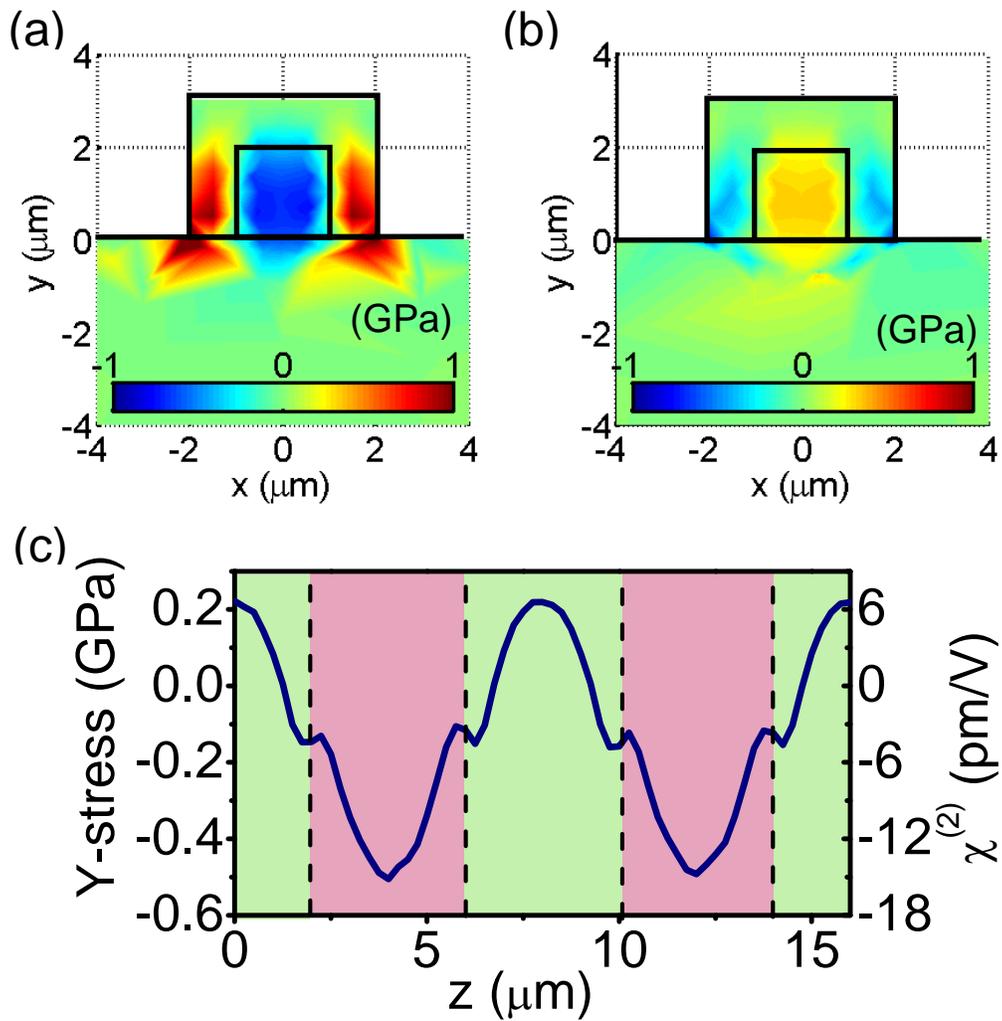

Fig. 2. (Color online) Simulated cross-sectional *y*-component normal stress ($\sigma_{yy}$) distribution where the PePSi waveguide is covered by silicon nitride films with 1 GPa (*tensile*) stress and (b) –500 MPa (*compressive*) stress. (c) Simulated average $\sigma_{yy}$ and estimated second order susceptibility ($\chi^{(2)}$) induced along the waveguide (z-direction).

It has been previously shown that the application of asymmetric stress can break this symmetry, thereby inducing a nonzero second-order nonlinear coefficient [4]. The applied forces induce alternating stress gradients, which in turn induce an alternating dipole moment in the medium. In our calculations, we utilize a value for the stress-induced nonlinear coefficient based on the measurements of R. S. Jacobsen, et al. [4]. We also make the reasonable assumption that the sign of the induced coefficient depends on the direction of applied stress.

Interestingly, a simple formula constructed from parameters of the classical anharmonic oscillator model produces a value for the nonlinear coefficient that qualitatively agrees with the measurements of Jacobsen et al. Based on a classical anharmonic oscillator model [12], the stress-induced $\chi^{(2)}$ can be estimated as

$$\chi^{(2)} = 4q^3 S /(m^2 \varepsilon \omega^4 a^4),$$ where $q$ is the electron charge, $m$ is the electron mass, $\varepsilon$ is the dielectric permittivity, $a$ is the lattice constant, $\omega$ is the angular frequency of light, and $S$ is the strain, which is related to the stress by Young's modulus of the material and an element of the stress tensor. Thus, based on the simulated stress values, the induced $\chi^{(2)}$ is estimated to have oscillatory values from –15 pm/V to +6 pm/V in a period of 8 μm (Fig. 2(c)). The estimated $\chi^{(2)}$ agrees qualitatively with the prior work for the same order of magnitude of stress [4]. We would like to note that this formula, included in our report, is meant as a heuristic tool, and should not be interpreted rigorously. The values measured by Jacobsen et al. are the fundamental justification for the value of the nonlinear parameter used in our simulations. A full calculation of the nonlinearity induced in silicon by applied stress remains an open question, but was not required for our proof-of-principle report.

The PePSi technology introduced here allows efficient silicon photonic devices based on different $\chi^{(2)}$ processes. As an example, we consider generation of MWIR radiation in PePSi waveguide from two NIR beams using DFG. We numerically investigated the QPM-DFG process in the PePSi waveguide using the nonlinear Schrödinger equation (NLSE), which incorporates (1) stress-induced $\chi^{(2)}$ effects, (2) $\chi^{(3)}$ effects (Kerr effect, TPA). We simultaneously calculate the free-carrier concentration resulting from TPA and utilize known empirical relations, proposed by Soref and Bennet, to determine the associated free-carrier absorption (FCA) and refraction (FCR) [13].

Taking the waveguide dispersion into account, the 8-µm-period of the SiN film pattern is designed to satisfy the QPM condition in the DFG process: pump at 1.3 µm, signal at 1.75 µm and idler at 5.1 µm. We consider the fundamental TE-polarized modes of the three interacting waves. Although the waveguide is multi-mode at the pump and signal wavelengths, the higher-order modes are expected to have no significant effect on the QPM-DFG process if the pump and signal propagate predominantly in the fundamental modes because higher-order modes have different mode profiles and dispersion properties, which lead to different phase-matching conditions.

In the model, we input transform-limited pump and signal pulses (both with pulse width of 12 ps), which have peak intensities of 1.5 GW/cm$^2$ and 12.5 MW/cm$^2$, respectively, into a 2-cm-long PePSi waveguide. The peak intensities are chosen to obtain the highest conversion efficiency. Through QPM-DFG, the MWIR idler pulse can be efficiently generated at 5.1 µm at the waveguide output (Fig. 3). During the process, second harmonic generation (SHG) and sum frequency generation (SFG), albeit without phase matching, can generate photons with energies above the silicon bandgap. This inevitably causes single-photon absorption (SPA), which introduces additional FCA and, thus, deteriorates the DFG efficiency. We have incorporated both SHG and SFG in the model, but found that the efficiencies of both processes are negligibly low along the waveguide because both second-harmonic and sum-frequency waves are highly attenuated by SPA in the early stage of propagation. On the other hand, since the second-harmonic and sum-frequency waves remain weak throughout the waveguide, SPA is insignificant. We find that the average free-carrier concentration along the waveguide generated by SPA (due to SHG and SFG) is ~ $10^{12}$ cm$^{-3}$, whereas that generated by TPA is ~ $10^{15}$ cm$^{-3}$. Hence, the contribution of SPA to the overall FCA is still overwhelmed by that of TPA. We also remark that even though the idler wave suffers from larger FCA than the pump and signal waves due to the wavelength-dependence of FCA, only the trailing edge of the idler pulse is suppressed by the FCA since more free carriers are generated toward the trailing edge [14]. As a result, high peak conversion efficiency can still be obtained.

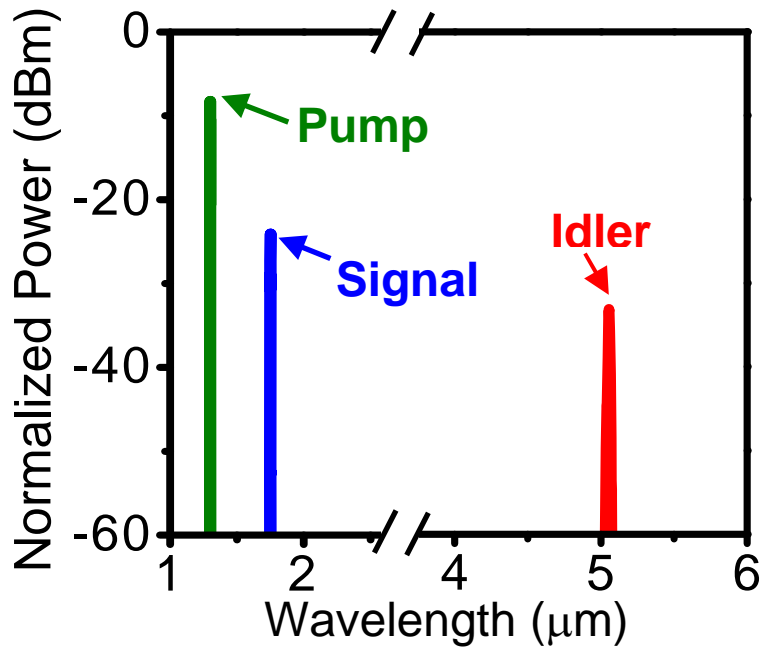

Fig. 3. (Color online) Calculated output spectra in a 2-cm-long PePSi waveguide (Period ~ 8 μm): a 12-ps pump pulse at 1.3 μm (input peak intensity = 1.5 GW/cm$^2$), a 12-ps signal pulse at 1.75 μm (input peak intensity = 12.5 MW/cm$^2$), idler at 5.1 μm.

As depicted in Fig. 4, which shows the peak conversion efficiency as a function of the peak pump intensity, the PePSi waveguide is able to achieve a maximum conversion efficiency of the MWIR generation by QPM-DFG as high as ~ –3 dB (~ 50%) under peak pump intensity of 1.5 GW/cm$^2$. This intensity level is readily achievable under experimental conditions. The slight decrease of conversion efficiency at high pump intensities observed in Fig. 4 is primarily due to TPA and FCA in the pump, signal, and idler waves, as well as FCR, which introduces additional phase-mismatch in the DFG process.

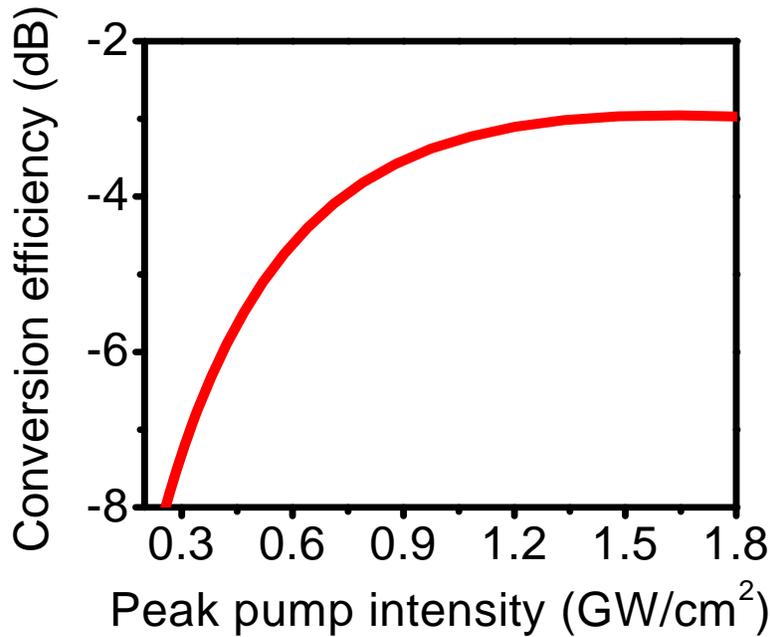

Fig. 4. (Color online) Calculated conversion efficiency of MWIR generation.

In summary, we have proposed a new type of photonic device based on periodically-poled silicon (PePSi). Our approach employs periodic stress fields in silicon waveguides such that the crystal symmetry of silicon is broken in a periodically alternating fashion. Introducing, what is in effect, the functional equivalent of periodic poling technology into silicon photonics offers a path to realize efficient wave mixing devices based on second-order optical nonlinearity. As an example of the utility of the PePSi technology, we numerically show that using QPM-DFG, MWIR generation at 5.1 μm in PePSi waveguide with conversion efficiency as high as ~ 50% can be achieved. When combined with piezoelectric stressed layers, this technology will offer the capability to dynamically control the $\chi^{(2)}$ in silicon.